# A geometric probabilistic approach to random packing of hard disks in a plane[1]

H.J.H. Brouwers
*Department of the Built Environment, Eindhoven University of Technology*
*P.O. Box 513, 5600 MB Eindhoven, The Netherlands*
*Email:jos.brouwers@tue.nl*
(Dated: November 8, 2023)

In this paper the random packing fraction of hard disks in a plane is analyzed, following a geometric probabilistic approach. First, the random close packing (RCP) of equally sized disks is modelled. Subsequently, following the same methodology, a simple, statistical geometric model is proposed for the random loose packing (RLP) of monodisperse disks. This very basic derivation of RLP leads to a packing value (≈ 0.66) that is in very good agreement with values that have been obtained previously for 2D disk packings. The present geometrical model also enables closed-form expression for the contact (coordination) number as function of the packing density at different states of compaction. These predictions are thoroughly compared with empirical and simulation results, among others the Rényi parking model, yielding good agreement.

## 1. INTRODUCTION

The packing of hard disks is of interest in mathematics and physics. In the mathematical branch it is generally known as "circles packing". In physics the packing of spheres on a plane, which is equivalent to that of circles in a plane, is studied as an introduction to the 3-dimensional (3D) problem. Furthermore, the packing of disks and spheres in a plane has been used to model the structure of monolayer films, the adsorption on substrates, and the organization of cells [1-3].

For ordered/regular packings in the 2-dimensional (2D) Euclidean plane, Lagrange proved in 1773 that the highest-density lattice packing of circles is the hexagonal arrangement, in which the centers of the circles are arranged in a triangular lattice, and each circle is surrounded by 6 other circles (Fig. 1a), packing fraction $\varphi_{tri} = \pi/2\sqrt{3} \approx 0.907$. This packing was proved to be the densest disk packing in the plane, so optimal among all packings, not just lattice packings, by Thue in 1910 [4, 5], although some consider Thue's proof flawed [2]. Fejes Tóth published a rigorous proof that this triangular packing has optimal packing indeed [2, 6]. In Fig. 1(b) also the honeycomb hexagonal lattice is shown, which can be seen as the union of two offset triangular lattices, with packing fraction $\varphi_{hon} = \pi/3\sqrt{3} \approx 0.605$.

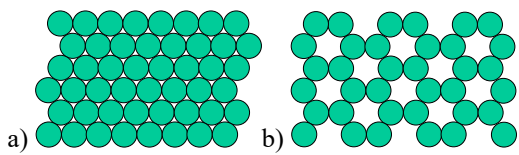

FIG. 1. Hexagonal disk lattices, (a) Triangular (b) Honeycomb.

Randomly, or jammed, close packed disks in 2D were first studied 60 years ago [7], where a random packing fraction of 0.89 $\varphi_{tri}$ (≈ 0.807) was measured. Experimental, modelling and simulation later studies yield $\varphi_{rcp} \approx 0.82$-$0.89$ [7-26]. Higher packing fractions, $\varphi_{rcp} \approx 0.88$-$0.89$, were found by Shahinpoor [16] and Zaccone [26], the latter's result leading to scientific debate [26]. Random loose packing (RLP) fraction studies have resulted in $\varphi_{rlp} \approx 0.66$-$0.84$ [3, 12, 19, 24, 27-30]. The lowest of these packing fraction values were reported recently [29-30].

In this paper, the random close packing (RCP) and random loose packing fraction (RLP) of hard disks is analyzed, following a geometric probabilistic approach. Statistical geometry (or integral geometry) can be traced back to the three problems formulated in the 18th century by George-Louis Leclerc, better known as Comte de Buffon. These three mathematical games of chances were the clean tile problem, the needle problem and the grid problem [31, 32].

Here, in Section 2 first a geometric probabilistic model for RCP is presented, an approach also followed in [10]. Next, in Section 3 a new model is put forward for RLP. Subsequently, in Section 4 the statistical models are used to obtain closed-form expressions between packing fraction and number of contacts (coordination number). In Section 5 the classic Rényi probability theory of car parking [33-35] is applied to assess and confirm the obtained RCP and RLP contact numbers. The conclusions are collected in Section 6.

For convenience, in this paper the diameter of the congruent disks (circles) is set to unity, so that the disk area is $\pi/4$ and the circumference $\pi$.

## 2. RANDOM CLOSE PACKING

---

[1] This article is dedicated to my late study friend ir. M.E. (Marc) van der Ree (26 December 1961 – 16 April 2000).

The packing of RCP has been studied experimentally, numerically and using analytical, relatively simple, models based on statistical geometry [3, 7-26]. In [3] a packing fraction of 0.824 for maximum random jammed (MRJ) disks was numerically generated, whereby the MRJ state can be thought of as the "most random" state of packing

For RCP, the method used is very similar to that reported in [10, 20], who considered a unit cell formed by 4 disks, the centers of the 4 disks forming a parallelogram, $\alpha$ being the angle between neighbor disks (Fig. 2), the enclosed disk area $A_d$ being $\pi/4$. For RCP, with $\alpha = \pi/3$, locally a triangular domain is obtained, considered as the densest possible packing configuration, and with $\alpha = \pi/2$ a square lattice, with $\varphi_{sq} = \pi/4$ ($\approx 0.785$).

The area of the parallelogram reads

$$A_t = \sin \alpha . \qquad (1)$$

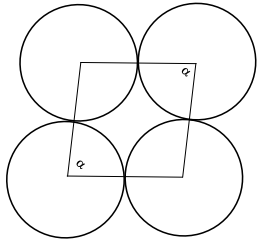

FIG. 2. Schematic representation of RCP unit.

And the packing fraction $\varphi$ follows as

$$\varphi = \frac{A_d}{A_t} = \frac{\pi}{4A_t} . \qquad (2)$$

Assuming a uniform probability density for the angle $\alpha$ between $\pi/3$ and $\pi/2$, the RCP density follows from Eqs. (1) and (2) as

$$\varphi_{rcp} = \frac{\pi \int_{\pi/3}^{\pi/2} (\sin x)^{-1} x dx}{4 \int_{\pi/3}^{\pi/2} dx} =$$

$$\frac{3}{2} \ln \left( \tan(\frac{x}{2}) \right) |_{\pi/3}^{\pi/2} = \frac{3}{4} \ln 3 \quad , \qquad (3)$$

yielding $\varphi_{rcp} \approx 0.824$.

This value is in accordance with the majority of measured and modeled results [1, 3, 7-15, 17-25]. This confirms that the statistical approach of the considered unit, with $\alpha$ uniformly distributed from $\pi/3$ to $\pi/2$, is representative for RCP. For $\alpha = \pi/3$ and $\alpha = \pi/2$ the ordered triangular and the square lattice domains, respectively, are obtained locally, and all other $\alpha$ values account for the disordered random nature.

An alternative and even more straightforward approach is based on the area pertaining to the mean angle $\alpha$ [20]. Using the arithmetic mean of $\alpha = \pi/3$ and $\alpha = \pi/2$, $\bar{\alpha} = 5\pi/12$, Eq. (1) yields $\bar{A}_t = 0.483$, and hence Eq. (2) $\varphi_{rcp} \approx 0.813$. Also this value is close to the aforesaid values.

## 3. RANDOM LOOSE PACKING

By introducing order [3] or friction [29, 30, 36, 37], a jammed random packing can be generated with lower packing fractions (and coordination numbers). Also unjammed or glass-like packings have a lower packing fraction. This lower random packing limit was explored by [12, 19, 24, 27-30]. Recently, from numerical simulations Pica Ciamarra and Coniglio [29] estimated $\varphi_{rlp} = 0.675$, Jin et al. [30] $\varphi_{rlp} = 0.67$, and Atkinson et al. [3] mapped $\varphi_{rlp} = 0.66$ as lowest jammed packing fraction.

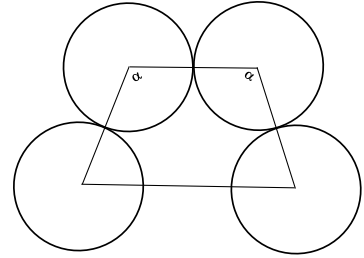

FIG. 3. Schematic representation of RLP unit.

As *ansatz* for a statistical approach of RLP, a unit cell of a quadrangle formed by 4 disks is proposed (Fig. 3), the assembly forming a trilateral trapezoid (or isosceles trapezoid). The assembly is governed by the angle $\alpha$, $\pi/2 \leq \alpha \leq 2\pi/3$. From elementary geometric considerations it follows that

$$A_t = \sin \alpha - \frac{\sin 2\alpha}{2} , \qquad (4)$$

and that the disk area $A_d$ of the 4 disks in the quadrangle is again $\pi/4$.

When $\alpha$ is $2\pi/3$, the presented RLP unit allows for a local honeycomb domain, and for $\alpha$ being $\pi/2$, a square lattice, similar as with the presented RCP unit (Fig. 2). Assuming a uniform probability density for the angle $\alpha$ between $\pi/2$ and $2\pi/3$, the RLP density follows from Eqs. (2) and (4) as

$$\varphi_{rlp} = \frac{\pi \int_{\pi/2}^{2\pi/3} \left( \sin x - \frac{\sin 2x}{2} \right)^{-1} dx}{4 \int_{\pi/2}^{2\pi/3} dx} , \qquad (5)$$

yielding $\varphi_{rlp} \approx 0.662$. This value is in the range of computer modeled values, *viz.* 0.66-0.675 [3, 29-30].

For completeness, also the RLP fraction is computed based on the arithmetic mean of $\alpha = \pi/2$ and $\alpha = 2\pi/3$, $\bar{\alpha} = 7\pi/12$, *i.e.* again following the approach by Williams [20].

Eq. (4) yields $\overline{A}_t \approx 1.216$, and hence from Eq. (2) it follows that $\varphi_{rlp} \approx 0.646$.

In other words, whichever method followed, a uniform probability density for the angle α or a mean angle $\overline{\alpha}$, the presented simple derivation produces a RLP packing value that is slightly different from those that have been obtained previously.

## 4. CONTACT NUMBER

The presented approach, in the previous sections, for RCP and RLP also allows correlating the packing fraction to the number of contact points (or coordination number) Z.

### 4.1 Present model

The coordination number Z can vary when the mean angle is lower or higher than the geometrically averaged ones determined in the previous sections. This can be the case when the state of MRJ packing has not yet been attained, when friction is introduced, ordering takes place, etc.

For the RCP model, with

$$Z = \frac{2\pi}{\alpha}, \qquad (6)$$

a mean coordination $Z = 4.8$ for mean $\overline{\alpha} = 5\pi/12$ is obtained. Alexander [38] demonstrated a minimum of $Z = 4$ for frictionless spheres. Assuming maximum disorder (and excluding local order), Bideau *et al.* confirmed experimentally and numerically that in such case the maximum coordination number is 4 [18], which may be lower when friction is at play. Also from packings studies it followed that random or MRJ packings are isostatic, so $Z = 4$ (2d, with d the dimension) for disks [3, 22, 23]. The present model is not able to reproduce this value, according to the model $Z = 4$ corresponds to the square lattice, with packing fraction $\varphi_{sq} = \pi/4$ ($\approx 0.79$).

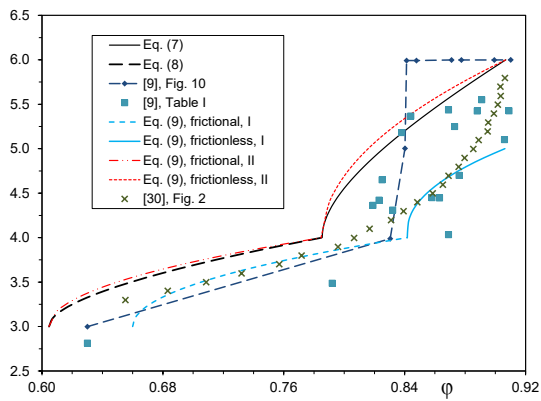

FIG. 4. Phase diagram of 2D disk packings, mean number of contacts Z versus packing fraction φ. Frictional I: $\zeta = 0.5$, $Z_0 = 2.35$, $Z_c = 3$ and $\varphi_c = 0.660$, frictionless I: $\zeta = 0.5$, $Z_0 = 3.55$, $Z_c = 4$ and $\varphi_c = 0.842$, frictional II: $\zeta = 0.5$, $Z_0 = 2.35$, $Z_c = 3$, and $\varphi_c = \varphi_{hon} = \pi/3\sqrt{3} \approx 0.605$, frictionless II: $\zeta = 0.5$, $Z_0 = 5.74$, $Z_c = 4$, and $\varphi_c = \varphi_{sq} = \pi/4 \approx 0.785$.

For all configurations $\pi/3 \leq \alpha \leq \pi/2$, so $4 \leq Z \leq 6$, φ can be expressed algebraically in Z by combining Eqs. (1), (3) and (6), yielding

$$\varphi = \frac{\pi}{4\sin\left(\frac{2\pi}{Z}\right)}. \qquad (7)$$

This relation is included in Fig. 4 for Z ranging from 4 to 6, and hence φ from $\pi/4$ to $\pi/2\sqrt{3}$.

For the RLP model, using mean $\overline{\alpha} = 7\pi/12$, Eq. (6) yields as mean coordination number $Z = 24/7 \approx 3.43$. This value is in line with Williams [27] and Uhler and Schilling [28] who found values of 3.2 and 3.333-3.416, respectively, by geometric probability calculations. From numerical simulations Hinrichsen *et al.* [19] assessed $Z = 3.02$. Note that $Z = 3 (d + 1)$ is the "loosest packing condition of Hilbert" [38]. Atkinson et al. [3] reported a tendency towards a kagomé (or trihexagonal) ordering ($\varphi_{trh} = \pi\sqrt{3}/8 \approx 0.68$) near their RLP packing fraction ($\varphi_{rlp} = 0.66$), having $Z = 4$.

Coordination number and packing fraction can also be related in closed-form in the range $3 \leq Z \leq 4$. By combining Eqs. (1), (4) and (6) it follows:

$$\varphi = \frac{\pi}{4\sin\left(\frac{2\pi}{Z}\right) + 2\sin\left(\frac{4\pi}{Z}\right)}, \qquad (8)$$

which is also set out in Fig. 4, for Z ranging from 3 to 4, and hence φ from $\pi/3\sqrt{3}$ to $\pi/4$.

### 4.2 Empirical data and other models

In Fig. 4 also the the experimental values of [9] taken from "Table I" and "Fig. 10" are included. The outlying packing fractions 0.090, 0.092, 0.098 and 0.385 of "Table I" [9] are omitted. The depicted values in Fig. 4 are tabulated in Tables A.1 and A.2 and (Appendix).

The scatter of the experimental points of "Table I" [9] is considerable, but the positive relation between coordination number and packing density becomes obvious. Quickenden and Tan [9] observed local ordering (crystallization), which became more pronounced with higher packing fractions. The points/line from "Fig. 10" [9] furthermore illustrates the sharp decrease in coordination number (from sixfold to fourfold) in a relatively small packing fraction change, which is captured by Eqs. (7) and (8).

By developing a "jamming phase diagram" based on numerical simulations, in [22, 23] for frictionless disks in 2D and spheres 3D a fit of the form:

$$Z - Z_c = Z_0(\varphi - \varphi_c)^\zeta, \qquad (9)$$

was derived, where ζ is independent of potential, polydispersity, and dimensionality d. For 2D, $\zeta \approx \frac{1}{2}$, $Z_c = 4$ (2d) and $\varphi_c = 0.842$. For 3D packing of spheres, the same ζ was fitted. The fact that ζ is independent of potential is intriguing because it suggests that ζ depends only on the geometry of the packing. The fact that ζ is also independent of dimensionality suggests that there is a property of the packing that is independent of dimension d [22, 23].

For frictionless monodisperse disks, $Z_0$ was not given and is therefore extracted from "Fig. 9" [22], see Table A.3 (Appendix), yielding $Z_c = 4$, $\varphi_c = 0.842$ and $Z_0 = 3.55$. As verification the same procedure is followed for the spheres given in "Fig. 9" [22], taking $Z_c = 6$ and $\varphi_c = 0.639$ [22, 23], and the provided value $Z_0 = 7.7$ could be reproduced.

In Fig. 4, Eq. (9) is set out for $Z_0 = 3.55$, $Z_c = 4$ and $\varphi_c = 0.842$, $\varphi$ ranging from 0.842 to $\varphi_{tri}$ (frictionless I). One can see that the equation yields $Z \approx 4.9$ for $\varphi_{tri}$ (instead of 6). This is not surprising as Eq. (9) was derived on for $(\varphi, Z)$ in the vicinity of $(\varphi_c, Z_c)$. If $Z_0$ where 7.5, the value of 3D, the right endpoint of the curve would be $(\varphi, Z) = (0.907, 6)$ instead of $(0.907, 4.9)$.

To examine $(\varphi, Z)$ for RLP of disks the paper by Silbert [38] is instrumental. In [38], $(\varphi, Z)$ of both frictional and frictionless spheres were simulated. In Tables A.4 and A.5 (Appendix) the extracted values for frictionless and frictional spheres packings, respectively, are listed, taken from "Fig. 9", they are shown in Fig. 5.

Also here it appears that Eq. (9) is applicable for both packings, again with $\zeta = \frac{1}{2}$ and $Z_0 = 7.7$ for both packing configurations. Note that in [22, 23] the same $\zeta$ and $Z_0$ where proposed for the 3D frictionless sphere packing, which values seem to be applicable to frictional spheres as well. For frictionless spheres it appears also that the values $Z_c = 6$ and $\varphi_c = 0.639$ are identical as in [22, 23]. In Fig. 5, Eq. (9) is set out with these values, and also for $Z_c = 4$ and $\varphi_c = 0.560$ for frictional spheres. The isostatic condition requires that for frictional spheres the critical $Z_c = d + 1$, and for frictionless spheres $Z_c = 2d$ [37]. One can see that with aforesaid values Eq. (9) follows that numerical data of [38] very well, both for frictional and frictionless spheres. So again Eq. (9) is consistent with a broad set of data.

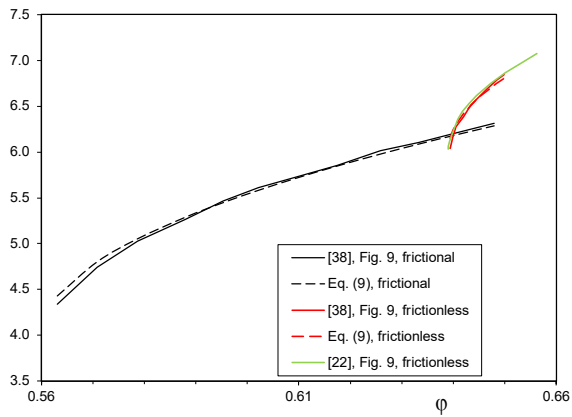

FIG. 5. Number of contacts Z versus packing fraction $\varphi$ for packing of spheres (3D). Eq. (9) frictional: $\zeta = 0.5$, $Z_0 = 7.7$, $Z_c = 4$ and $\varphi_c = 0.56$, Eq. (9) frictionless: $\zeta = 0.5$, $Z_0 = 7.7$, $Z_c = 6$ and $\varphi_c = 0.639$.

Eq. (9) is apparently applicable to both frictional and frictionless spheres, so all $(\varphi, Z)$, even with the same $\zeta$, and this also holds for frictionless disks. It is therefore plausible to propose that Eq. (9) is also applicable to frictional disks, so in the lower $(\varphi, Z)$ range of Fig. 4, again with $\zeta = 0.5$, $Z_c = 3$ (= d + 1) and $\varphi_c = 0.66$ (frictional I). By invoking $Z_0 = 2.35$, Eq. (9) has as right endpoint $(\varphi, Z) = (0.842, 4)$, which is the left endpoint of the frictionless disk packing curve (Fig. 4).

The frictionless and frictional sphere curves of [38] are concave and have a discontinuity at $Z = 2d$ (Fig. 5), which is also the case for Eqs. (7) and (8). Only the packing fraction values at $Z = 3$, 4 and 6 are not equal. To illustrate the similarity of Eqs. (7)-(9), In Fig. 4 Eq. (9) is set out for $\zeta = 0.5$, $Z_0 = 2.35$, $Z_c = 3$, and $\varphi_c = \varphi_{hon} = \pi/3\sqrt{3} \approx 0.605$, and for $\zeta = 0.5$, $Z_0 = 5.74$, $Z_c = 4$, and $\varphi_c = \varphi_{sq} = \pi/4 \approx 0.785$ (frictional II and frictionless II, respectively). Note that for $\varphi = \pi/3\sqrt{3}$, $\pi/4$ and $\pi/2\sqrt{3}$, these equations yield $Z = 3$, 4 and 6, respectively. From Fig. 4 it follows that Eqs. (7) and (8), based on the simple geometric model, are following Eq. (9) quite closely. This latter correlation has its foundation in the results of thermodynamic modelling.

All data in Fig. 4 feature an increase of Z with packing fraction $\varphi$, as expected. But following the presented models and [22, 23, 38], $Z(\varphi)$ is a concave function in 2D and 3D. Also the experiments by [9] suggest a concave-shaped relation. For [22, 23, 38] this concave trend follows from the exponent $\zeta \approx \frac{1}{2}$, which is smaller than unity.

In Fig. 4 also $(Z, \varphi)$ is included as computed by Jin et al. [30] pertaining to $\varphi_{RCP} = 0.804$ ("n = 3"), listed in Table A.6 (Appendix). The data shows a convex $Z(\varphi)$ relation, so a power $\zeta > 1$. Also the model by Zaccone et al. [26] and Anzivino et al. [39] yields a convex equation. This can be attributed to their assumption that Z is proportional to the product of $\varphi$ and the semi-empirical Carnahan-Starling (CS) expression. For d = 2, this CS expression reads $(1 – 0.436\varphi)/(1 - \varphi)^2$ [26].

In all, the phase diagram of the 2D disk packing shows that Eqs. (7) and (8) are able to correlate reported measured and simulated $(Z, \varphi)$ quite well, especially considering the simplicity of the presented model.

## 5. PARKING PROBLEM APPROACH

Rényi's [33, 34] classic parking problem (or 1D sequential interval packing problem) concerns the probabilistic properties of the following random process: consider an interval of length x ($x \geq 0$), with x eventually tending to infinity, and sequentially and randomly pack disjoint unit in x as long as the remaining space permits placing any new unit segment. At each step of the packing process the position of the newly placed interval is chosen uniformly from the available space.

The expected value of the covered part is denoted by M(x), so that the ratio M(x)/x is the expected filling density of the "parking process". The interval x is the street curb, and the packed unit segments are the parked cars. The placing of disks on a center disk can be modelled using this parking model (Fig. 6). The circumference of the center disk corresponds to 6 units (disks) and can be seen as a circular street curb. After parking the first unit on a center sphere, a closed interval of length $x = 5$ is remaining in which the remaining units (disks) are randomly placed on the center sphere.

Weiner [35] presented the following lower and upper limits for $M(x)$ for $x > 4$:

$$0.7432\,x - 0.2568 \leq M(x) \leq 0.75\,x - 0.25, \quad (10)$$

yielding $3.46 \leq M(5) \leq 3.5$.

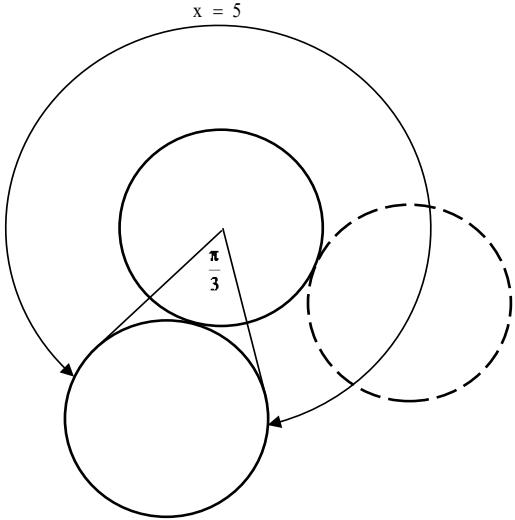

FIG. 6. Schematic representation of randomly placing ("parking") spheres on a center sphere. After placing the first sphere (solid line) there is place left for the parking of maximum 5 other spheres ("cars"), *i.e.,* the "street interval" length $x = 5$. The dotted line sphere represents the first randomly placed sphere.

These values are compatible with the approximate solution provided by Rényi [33, 34] for large x:

$$M(x) \approx Cx - (1 - C), \quad (11)$$

with $C \approx 0.748$, yielding $M(5) \approx 3.49$. C is the Rényi parking constant which is the asymptotic mean filling density $M(x)/x$ for $x \to \infty$. To summarize, the parking approach to the RCP disk packing yields $Z \approx 4.48$, which is not far from the value 4.8 derived here.

The parking problem approach can also be applied to RLP of disks. To allow for obtaining a lower parking density, each unit (disk) that is placed on the center disk is considered to have extra excluded space in which another unit cannot be parked. With the estimated $Z \approx 3.35$ for RLP, the space taken by one unit can be estimated.

After parking of the first unit, in the remaining interval x, 2.35 units can be parked. The following equation [34]

$$M(x) = 7 - \frac{10}{x-1} - \frac{4\ln(x-2)}{x-1}, \quad (12)$$

is exact for $3 \leq x < 4$, and with $M(x) = 2.35$, it follows that $x = 3.5$. The total interval length x is hence 4.5 units, *i.e.* the circumference of the center disk can accommodate 4.5 loosely placed disks. In other words, where in case of random close packing the circumference can accommodate 6 disks, in random loose packing it is about 75% of this value. So, in RCP each disk occupies $\pi/6$ (or $3\pi/18$) of the center disk's circumference of $\pi$, as the disk diameter is unity (Fig. 5). For RLP, on the other hand, the excluded space of each disk effectively is about $4\pi/18$ of this circumference.

## 6. CONCLUSIONS

This paper addresses the ancient problem of random packing of hard disks/circles in a plane, of interest in mathematics science and engineering, following a geometric probabilistic approach.

Simple models for RCP and RLP units are constructed (Figs. 2 and 3), which assume that each state can be visited with equal probability. They result in packing fractions and contact numbers $\varphi_{rcp} = 3\ln(3)/4 \approx 0.824$ and $Z = 4.8$ for RCP, and $\varphi_{rlo} \approx 0.662$ and $Z = 24/7 \approx 3.43$ for RLP. Both Z-values are higher than the minimum Z required by isostatisticy, $Z = 4$ (2d) and $Z = 3$ (d +1), respectively. The derived packing fraction values are commonly observed, in other probabilistic studies, more sophisticated computer simulations, and experimental studies.

The derived mean contact number of RCP, $Z = 4.8$, is also very close to the number that is obtained here by applying the classic Rényi parking theory, viz. $Z \approx 4.5$. Applying the Rényi parking solution to $Z \approx 3.35$ for RLP, yields that the disks occupy $4\pi/18$ of the center disk's circumference of $\pi$ on which they are placed, whereas it is $\pi/6$ for RCP (resulting in aforesaid $Z = 4.8$).

It also appears that the present approach allows for a phase diagram of average contact number versus 2D disks packing fraction (Fig. 4), based on closed-form expression Eqs. (7) and (8). Also these equations produce results that are only slightly different from Eq. (9) provided by [22, 23] both for packed disks in 2D and for spheres in 3D.

In line with the geometric probabilistic approach [10, 11, 31] and Rényi's parking model [33, 34], a uniform distribution of the disks is invoked here. In other words, there is no preference for α (Figs. 2 and 3) nor where a car/sphere is placed (Fig. 6). In future work it would be interesting to study the effect of non-uniform distributions (*e.g.* by perturbing the uniform distribution) on the packing fraction and mean contact number.

### ACKNOWLEDGEMENTS

Dr. Lucas Gerin from CMAP, École Polytechnique, Palaiseau (France) is acknowledged for providing and explaining [35]. Dr. Anna Kaja is thanked for retrieving the graphical data of [9, 30], see Tables A.2 and A6, and

**Appendix**

| φ | Z | φ | Z | φ | Z |
|---|---|---|---|---|---|
| 0.630 | 2.999 | 0.841 | 5.992 | 0.878 | 5.999 |
| 0.830 | 3.988 | 0.848 | 5.992 | 0.899 | 5.999 |
| 0.840 | 5.004 | 0.871 | 5.999 | 0.910 | 5.999 |

Table A.1. Packing fraction φ versus coordination number Z, taken from "Fig. 10" of Quickenden and Tan [9].

| φ | Z | φ | Z | φ | Z |
|---|---|---|---|---|---|
| 0.832 | 4.31 | 0.876 | 4.70 | 0.909 | 5.43 |
| 0.63 | 2.81 | 0.906 | 5.10 | 0.792 | 3.49 |
| 0.869 | 4.03 | 0.844 | 5.37 | 0.819 | 4.36 |
| 0.863 | 4.45 | 0.873 | 5.25 | 0.858 | 4.45 |
| 0.823 | 4.42 | 0.888 | 5.43 | 0.838 | 5.18 |
| 0.825 | 4.65 | 0.891 | 5.55 | 0.869 | 5.44 |

Table A.2. Packing fraction φ versus coordination number Z, taken from "Table I" of Quickenden and Tan [9].

| log(φ − φ$_c$) | φ | log(Z − Z$_c$) | Z | Z$_0$ |
|---|---|---|---|---|
| -4.793 | 0.842 | -1.847 | 4.01 | 3.5 |
| - 4.203 | 0.842 | -1.559 | 4.03 | 3.5 |
| - 3.600 | 0.842 | -1.249 | 4.06 | 3.6 |
| - 3.001 | 0.843 | -0.950 | 4.11 | 3.6 |
| -2.398 | 0.846 | -0.651 | 4.22 | 3.5 |
| - 1.795 | 0.858 | -0.342 | 4.46 | 3.6 |

Table A3. Z$_0$ computed from packing fraction φ and coordination number Z, extracted from the disks (2D) line shown in "Fig. 9" of O'Hern *et al.* [22], using Eq. (9) and ζ = ½, Z$_c$ = 4, and φ$_c$ = 0.824.

| φ | Z | φ | Z | φ | Z |
|---|---|---|---|---|---|
| 0.563 | 4.34 | 0.595 | 5.46 | 0.626 | 6.01 |
| 0.571 | 4.74 | 0.602 | 5.61 | 0.633 | 6.10 |
| 0.579 | 5.02 | 0.610 | 5.73 | 0.641 | 6.21 |
| 0.588 | 5.27 | 0.617 | 5.86 | 0.648 | 6.31 |

Table A4. Packing fraction φ and coordination number Z, extracted from the frictional spheres (3D) as shown in "Fig. 9" of Silbert [38].

| φ | Z | φ | Z | φ | Z |
|---|---|---|---|---|---|
| 0.639 | 6.04 | 0.642 | 6.39 | 0.647 | 6.69 |
| 0.640 | 6.15 | 0.643 | 6.51 | 0.648 | 6.79 |
| 0.641 | 6.27 | 0.645 | 6.59 | 0.650 | 6.84 |

Table A5. Packing fraction φ and coordination number Z, extracted from the frictionless spheres (3D) as shown in "Fig. 9" of Silbert [38].

| φ | Z | φ | Z | φ | Z |
|---|---|---|---|---|---|
| 0.655 | 3.298 | 0.831 | 4.197 | 0.889 | 5.098 |
| 0.683 | 3.402 | 0.839 | 4.298 | 0.895 | 5.195 |
| 0.709 | 3.499 | 0.848 | 4.398 | 0.895 | 5.296 |
| 0.732 | 3.599 | 0.859 | 4.499 | 0.899 | 5.397 |
| 0.757 | 3.703 | 0.865 | 4.596 | 0.902 | 5.497 |
| 0.772 | 3.799 | 0.869 | 4.696 | 0.904 | 5.594 |
| 0.796 | 3.896 | 0.876 | 4.797 | 0.903 | 5.698 |
| 0.806 | 3.997 | 0.880 | 4.897 | 0.906 | 5.795 |
| 0.817 | 4.101 | 0.885 | 4.998 | | |

Table A6. Packing fraction φ versus coordination number Z, taken from "Figure 2 (n = 3)" of Jin *et al.* [30].